\documentclass[preprint,12pt]{aastex}
\usepackage{natbib} 
\begin{document}

\doublespace

\title{Comment on ``Chandra Reveals X-ray Glint in the Cat's Eye''}

\author{Holly L. Maness\altaffilmark{1,2}} 
\author{S. D. Vrtilek\altaffilmark{2}}

\altaffiltext{1}{Department of Physics, Grinnell College, Grinnell, IA 50112} 
\altaffiltext{2}{Harvard-Smithsonian Center for Astrophysics, Cambridge, MA 02138}

\begin{abstract}

\citet{Chu01} report that they find the diffuse X-ray emission from 
the Cat's Eye Nebula, NGC 6543, 
to be inconsistent with nebular abundances but consistent with
abundances expected for the stellar wind although the location and the 
derived temperature for the X-ray emitting
region of NGC 6543 require that a significant fraction of the
X-ray emitting material be nebular.  
A recent determination by \citet{BS03} of nebular
abundances in NGC 6543 incorporating infrared,
optical, and ultraviolet observations provides 
a resolution of this 
apparent anomaly.

\end{abstract}

\keywords {planetary nebulae: general--planetary nebulae: 
individual (NGC 6543)--stars: winds, outflows--X-rays: ISM}

\clearpage

\citet{Chu01} found that the X-ray spectrum of NGC 6543 
could be adequately fit using stellar wind abundances but not with 
abundances that had been reported for the nebular material in this
object \citep{Aller83, Pwa84,
Manchado89, 
deKoter96}.
\citet{Chu01} emphasized that the low 
temperature they derive
for the X-ray emitting material (1.7$\times 10^6$K) is 
puzzling since 
the expected postshock temperature for the fast (1750 km s$^{-1}$)
stellar wind of NGC 6543 is of the order of 10$^8$K. 
They also note that the location of the X-ray emitting gas
requires that a significant fraction of the X-ray emitting material be
nebular.
Here we re-analyze the {\it Chandra} X-ray Observatory ACIS-S spectrum
of NGC 6543 using updated abundances found for the nebular 
material in an attempt
to resolve this apparent anomaly.

Consistent with the work of \citet{Chu01} we extracted spectra 
from four regions:  the entire nebula and three
regions corresponding to the central elliptical
shell, the northern extension, and the southern extension.
We are 
able to reproduce 
the spectral results of 
\citet{Chu01}, finding that the stellar wind abundances they used 
modeled the
observed X-ray spectrum much better than the abundances they
used for the nebula.  

However, recently  
\citet{BS03} re-determined the nebular
abundances in NGC 6543 using
multiple observations in the infrared, optical, and ultraviolet.  
They find a N/O abundance ratio that is significantly increased 
from that of previous measurements and used by \citet{Chu01}.
Here we use the \citet{BS03} abundances  
in the 
VMEKAL (variable abundance thin plasma 
emission) model in the X-ray analysis package XSPEC \citep{AR96}. 
These are listed in Table 1 as abundances relative to Solar using
\citet{Grevesse98} values for Solar abundance. 
All elements fit by VMEKAL that are not determined by \citet{BS03} are
kept at
Solar abundances.  
We fit this new nebular model  
with only temperature, absorption column density, 
and the normalization factor as free parameters to the X-ray spectrum
extracted from the entire nebula.  The fits  
produced good results, giving significantly 
lower $\chi^2$ ($\chi_{\nu}^2$=1.3) than
both the nebular ($\chi_{\nu}^2$=1.8) and the stellar 
wind ($\chi_{\nu}^2$=1.6) models proposed by
\citet{Chu01}.  Allowing important abundances to vary does
not significantly lower $\chi^2$.  This new nebular model 
(plotted with the
spectrum extracted from the entire nebula in Figure 1a) gives a plasma 
temperature of {\it T} = 1.8 $\times$ 10$^{6}$ K and an absorption 
column density of {\it N}$_H$ = 2.5 $\times$ 10$^{20}$ cm$^{-2}$. 
The reduced $\chi^2$ as a function of {\it N}$_H$ and {\it kT} is 
plotted in Figure 1b.  The 99\% confidence contour
 spans {\it N}$_H$ = (2.0-3.5) $\times$ 10$^{20}$ cm$^{-2}$ and {\it T}
= (1.54-1.66) $\times$ 10$^{6}$ K.  For the energy range 0.2-1.5 keV
we obtain 
an absorbed X-ray flux of 9.1 $\times$ 10$^{-14}$ ergs 
cm$^{-2}$ s$^{-1}$ and an unabsorbed flux of 1.3 $\times$ 10$^{-13}$ ergs 
cm$^{-2}$ s$^{-1}$.  Assuming a distance of 1$\pm0.3$ kpc \citep{Cahn92,Reed99},
this gives a 0.2-1.5 keV X-ray luminosity of 
1.5 $\times$ 10$^{31}$ ergs s$^{-1}$.   

Fits to the spectra extracted from the central shell
and the northern and southern extensions using the \citet{BS03}
abundances 
do not produce significantly different results from those reported
earlier: 
consistent with the work of \citet{Miranda92}
and \citet{Chu01}, 
our fits suggest that temperature appears uniform across the 
X-ray emitting region of NGC
6543 with no significant variations detected whereas 
the intervening absorption is 
greater for the southern extension ({\it N}$_H$ = 4.3$\pm.02\times$
10$^{20}$ cm$^{-2}$) and the central shell 
({\it N}$_H$ = 3.7$\pm.02\times$ 10$^{20}$ cm$^{-2}$) than 
for the northern extension ({\it N}$_H$ = 1.2$\pm.08\times$ 10$^{20}$ cm$^{-2}$).  

We would like to thank J.Bernard-Salas, S.R.Pottasch, P.R.Wesselius 
and W.A.Feibelman for providing us their results prior to publication.
H.M. was supported by the NSF REU program at SAO.
SDV was supported in part by NASA grant NAG5-6711.

\begin{deluxetable}{lccc}
\tabletypesize{\scriptsize}
\tablecolumns{4}
\tablewidth{0pt}
\tablecaption{Abundances (relative to Solar) used to fit X-ray Spectra}
\tablehead{\colhead{}&\colhead{}&\colhead{This paper}}
\tablehead{\colhead{Element}&\colhead{Chu Nebular}
&\colhead{Chu Stellar}&\colhead{Bernard-Salas Nebular}}
\startdata
He&1.13&60&1\\
C&0.63&1&0.18\\
N&0.53&3&3.97\\
O&0.66&1&0.7\\
Ne&1.14&1&1.25\\
S&1&1&0.45\\
Ar&1&1&1.24\\
\enddata     
\end{deluxetable}

\clearpage
\begin{figure}
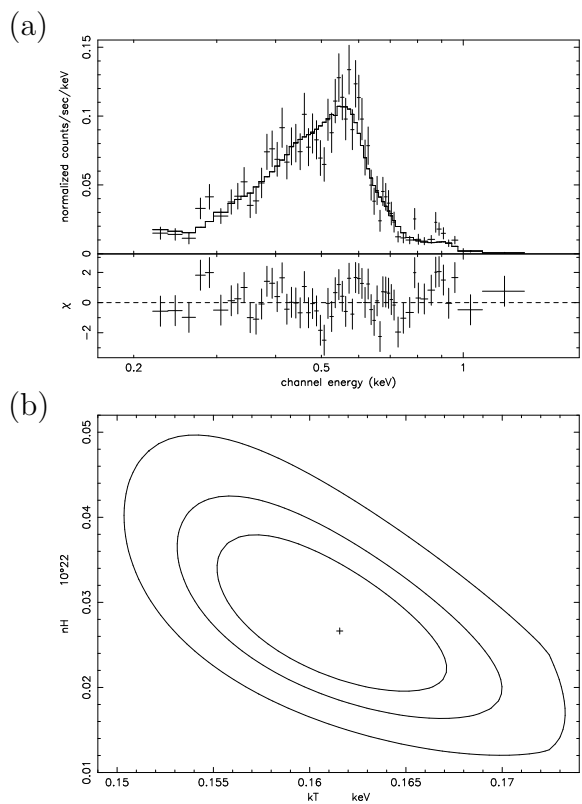

\centering
(a)
\scalebox{0.3}{\rotatebox{-90}{\includegraphics{f1.eps}}}

(b)
\scalebox{0.3}{\rotatebox{-90}{\includegraphics{f2.eps}}}
\caption{(a) X-ray spectrum of NGC 6543 with new nebular model
overlaid. (b) $\chi^2$ grid plots of nebular model spectral fit.  The
contours represent 68$\%$ (1 $\sigma$), 90$\%$ (2 $\sigma$), and 99$\%$ (3
$\sigma$) confidence levels. \label{new_6543}}
\end{figure}


\begin{thebibliography}{}  
\bibitem[Aller \& Czyzak(1983)]{Aller83} Aller, L.~H.~\& 
Czyzak, S.~J.\ 1983, \apjs, 51, 211
\bibitem[Arnaud (1996)]{AR96} Arnaud, K.A., 1996, in Astronomical Data 
Analysis Software and Systems V, eds. Jacoby G. and Barnes J., ASP 
Conf. Series volume 101,17
\bibitem[Bernard-Salas et al.(2003)]{BS03} Bernard-Salas, J., 
Pottasch, S.R., Wesselius, P.R.,
\& Feibelman, W.A. \ 2003, A\&A, submitted. 
\bibitem[Balucinska-Church \& McCammon(1992)]{Balucinska92} 
Balucinska-Church, M.~\& McCammon, D.\ 1992, \apj, 400, 699 
\bibitem[Cahn, Kaler, \& Stanghellini(1992)]{Cahn92} Cahn, 
J.~H., Kaler, J.~B., \& Stanghellini, L.\ 1992, \aaps, 94, 399  
\bibitem[Chu et al.(2001)]{Chu01} Chu, Y., Guerrero, M.~;., 
Gruendl, R.~A., Williams, R.~M., \& Kaler, J.~B.\ 2001, \apjl, 553, L69
\bibitem[Grevesse \& Sauval(1998)]{Grevesse98} Grevesse, N.~\&  
Sauval, A.J.\ 1998, Space Science Reviews, 85, 161
\bibitem[de Koter, Hubeny, Heap, \& Lanz(1996)]{deKoter96} de 
Koter, A., Hubeny, I., Heap, S.~R., \& Lanz, T.\ 1996, ASP Conf.~Ser.~96: 
Hydrogen Deficient Stars, 141 
Sauval, A.~J.\ 1998, Space Science Reviews, 85, 161  
\bibitem[Manchado \& Pottasch(1989)]{Manchado89} Manchado, A.~\& 
Pottasch, S.~R.\ 1989, \aap, 222, 219
\bibitem[Miranda \& Solf(1992)]{Miranda92} Miranda, L.~F.~\& 
Solf, J.\ 1992, \aap, 260, 397 
\bibitem[Pwa, Pottasch, \& Mo(1984)]{Pwa84} Pwa, T.~H., 
Pottasch, S.~R., \& Mo, J.~E.\ 1984, \aap, 139, L1  
\bibitem[Raymond \& Smith(1977)]{Raymond77} Raymond, J.~C.~\& 
Smith, B.~W.\ 1977, \apjs, 35, 419 
\bibitem[Reed et al.(1999)]{Reed99} Reed, D.~S., Balick, B., 
Hajian, A.~R., Klayton, T.~L., Giovanardi, S., Casertano, S., Panagia, N., 
\& Terzian, Y.\ 1999, \aj, 118, 2430 
\end{thebibliography}
\end{document}